\documentclass[twocolumn,english,prl,preprintnumbers,superscriptaddress]{revtex4-1}
\usepackage[utf8]{inputenc}
\usepackage{amsmath}
\usepackage{amssymb}
\usepackage{graphicx}
\usepackage{textcomp}
\usepackage{bm}
\usepackage[right]{eurosym}
\usepackage{float}
\usepackage[english]{babel}
\usepackage{blindtext}
\usepackage{babel}
\usepackage{booktabs}
\usepackage{longtable}
\usepackage{xcolor}

\begin{document}
\author{R. Michiels}
\affiliation{Institute of Physics, University of Freiburg, 79104 Freiburg, Germany}
\author{M. Abu-samha}
\affiliation{College of Engineering and Technology, American University of the Middle East, Kuwait}
\author{L. B. Madsen}
\affiliation{Department of Physics and Astronomy, Aarhus University, 8000 Aarhus C, Denmark}
\author{M. Binz}
\affiliation{Institute of Physics, University of Freiburg, 79104 Freiburg, Germany}
\author{U. Bangert}
\affiliation{Institute of Physics, University of Freiburg, 79104 Freiburg, Germany}
\author{L. Bruder}
\affiliation{Institute of Physics, University of Freiburg, 79104 Freiburg, Germany}
\author{R. Duim}
\affiliation{Institute of Physics, University of Freiburg, 79104 Freiburg, Germany}
\author{A. Wituschek}
\affiliation{Institute of Physics, University of Freiburg, 79104 Freiburg, Germany}
\author{A. C. LaForge}
\affiliation{Department of Physics, University of Connecticut, Storrs, Connecticut, 06269, USA}
\author{R. J. Squibb}
\affiliation{Department of Physics, University of Gothenburg, Sweden}
\author{R. Feifel}
\affiliation{Department of Physics, University of Gothenburg, Sweden}
\author{C. Callegari}
\affiliation{Elettra-Sincrotrone Trieste, 34149 Basovizza, Trieste, Italy}
\author{M. Di Fraia}
\affiliation{Elettra-Sincrotrone Trieste, 34149 Basovizza, Trieste, Italy}
\author{M. Danailov}
\affiliation{Elettra-Sincrotrone Trieste, 34149 Basovizza, Trieste, Italy}
\author{M. Manfredda}
\affiliation{Elettra-Sincrotrone Trieste, 34149 Basovizza, Trieste, Italy}
\author{O. Plekan}
\affiliation{Elettra-Sincrotrone Trieste, 34149 Basovizza, Trieste, Italy}
\author{K. C. Prince}
\affiliation{Elettra-Sincrotrone Trieste, 34149 Basovizza, Trieste, Italy}
\author{P. Rebernik}
\affiliation{Elettra-Sincrotrone Trieste, 34149 Basovizza, Trieste, Italy}
\author{M. Zangrando}
\affiliation{Elettra-Sincrotrone Trieste, 34149 Basovizza, Trieste, Italy}
\affiliation{CNR-IOM,~Elettra-Sincrotrone Trieste S.C.p.A., Italy }
\author{F. Stienkemeier}
\affiliation{Institute of Physics, University of Freiburg, 79104 Freiburg, Germany}
\author{M. Mudrich}
\affiliation{Department of Physics and Astronomy, Aarhus University, 8000 Aarhus C, Denmark}

\title[An \textsf{achemso} demo]
  {Enhancement of above threshold ionization in resonantly excited helium nanodroplets}
\begin{abstract}
Clusters and nanodroplets hold the promise of enhancing high-order nonlinear optical effects due to their high local density. However, only moderate enhancement has been demonstrated to date. Here, we report the observation of energetic electrons generated by above-threshold ionization (ATI) of helium (He) nanodroplets which are resonantly excited by ultrashort extreme ultraviolet (XUV) free-electron laser pulses and subsequently ionized by near-infrared (NIR) or near-ultraviolet (UV) pulses. The electron emission due to high-order ATI is enhanced by several orders of magnitude compared to He atoms. The crucial dependence of the ATI intensities with the number of excitations in the droplets suggests a local collective enhancement effect. 

\end{abstract}
\date{\today}
\maketitle

The nonlinear interaction of intense light with matter gives rise to stunning phenomena such as non-sequential double ionization~\cite{Watson:1997}, the emission of XUV and x-ray radiation by high-order harmonic generation (HHG)~\cite{Eberly:1991,Ganeev:2012}, and the acceleration of electrons and ions to high energies~\cite{Malka:2002,Schwoerer:2006}. In particular, HHG is widely used today for time-resolved XUV spectroscopy and attosecond science~\cite{Krausz:2009}. However, the conversion efficiency of HHG, usually performed in atomic gases, is notoriously low ($\lesssim10^{-5}$). Therefore, condensed-phase targets are being explored in view of enhancing the yield and photon energy of the radiation generated by HHG.

A closely related phenomenon is the emission of electrons with kinetic energies equal to multiples of the photon energy, termed above-threshold ionization (ATI)~\cite{Eberly:1991}. ATI occurs when an electron absorbs more than the minimum number of photons required for ionization and manifests itself by multiple equidistant peaks in electron spectra spaced by the photon energy. Corkum and Kulander developed a three-step semiclassical model which provides an intuitive understanding of the process leading to ATI and establishes the connection between ATI and HHG~\cite{Kulander:1993,Corkum:1993}. In this model, the radiation field lowers the potential barrier for an electron bound to an atom such that the electron may tunnel ionize. The free electron is then accelerated by the external field, and returns to the ion when the field reverses its direction. When the electron recollides with the ion, it scatters either elastically or inelastically. The latter gives rise to HHG, whereas in the former case, dephasing of the electron oscillation and energy absorption from the laser field further raise the electron energy $E_e$ leading to ATI. This model predicts a cutoff for the energy of emitted electrons at 10~$U_p$, where $U_p$ is the electron's ponderomotive energy~\cite{Paulus:1994}. 

Previous theoretical and experimental investigations of ATI have primarily focused on atomic targets. Molecules and clusters have additional degrees of freedom such as vibration, which was found to decrease the peak separation in the ATI electron spectra~\cite{Verschuur:1989} and induce multiple cutoffs up to 50~$U_p$~\cite{Chirilua:2006}. 
Using clusters, higher cutoffs were predicted as well~\cite{Faria:2002,Nguyen:2004}. The enhanced nonlinear response was discussed in the context of electron scattering from multiple centers~\cite{Moreno:1994,Bandrauk:1997,Veniard:1999,Vazquez:2001,Chirilua:2006}. Recently, electron-energy cutoffs far beyond 10~$U_p$ were found for argon clusters irradiated by intense ($>10^{14}$~Wcm$^{-2}$) NIR and mid-infrared pulses~\cite{wang2020universal}. An extended rescattering model taking into account the extended potential of a multiply ionized cluster reproduced the observed scaling of cutoff energies with cluster size and the quiver amplitude of the electron in the laser field, $x_0$. Alternatively, a macroscopic dipole moment resulting from collective oscillations of electrons about the cluster ions may lead to an enhancement of $E_e$ as observed in plasmonic nanostructures~\cite{Tisch:1997,Zherebtsov:2011,Passig:2017}. Likewise~\cite{Fu:2001}, clusters are promising systems for boosting HHG to higher photon energies and yields as they combine the advantage of a high local density of solids with the low average density of gases, making them dense yet transparent, renewable targets~\cite{Donnelly:1996,Tisch:1997,Vazquez:2001,Park:2014}. 

In this Letter we explore the nonlinear optical response of He nanodroplets prepared in multiply excited states by irradiation with XUV pulses. The excited nanodroplets are probed by
NIR (800~nm) and UV (400~nm) laser pulses at moderate intensities ($\leq 3\times 10^{13}$~Wcm$^{-2}$) where only the excited He atoms are ionized, whereas the ground state He atoms in the droplets remain inactive. We find drastically extended ATI structures in the electron spectra as compared to excited He atoms in the gas phase, both for NIR and UV probe pulses. A simple semi-empirical model for the collective enhancement of $E_e$ is presented. Similar ATI spectra generated by He$^+$ ions in excited states were recently observed for strong-field ionized He nanodroplets and were used to monitor the time evolution of the nanoplasma mean-field potential~\cite{kelbg2020temporal}. In He nanodroplets doped by single atoms or molecules of a different species, enhanced ATI-like electron structures were attributed to laser-assisted electron scattering upon the neutral He atoms surrounding the dopant~\cite{treiber2021observation}. In contrast, in the present experiment we show that a collective behavior of the active atoms plays a decisive role.


In this experiment, a pulsed jet of He nanodroplets was irradiated by XUV pulses (23.7~eV) generated by the seeded free-electron laser (FEL) FERMI in Trieste, Italy~\cite{Allaria:2012}. At this photon energy, both the free He atoms and the He droplets are resonantly excited into the $1s4p$ state. The FEL pulses (FWHM duration 70~fs, pulse energy at the endstation 2-50~nJ) were focused to a FWHM spot size of 70~$\mu$m~\cite{raimondi2019kirkpatrick}. Unless explicitly varied in the measurement, the intensity of the XUV pump pulses was $I_\mathrm{XUV} =1.8\times 10^{10}$~Wcm$^{-2}$. The probe pulses for ionization of the excited droplets were generated by a Ti:Sa laser (FWHM duration 100~fs) that was synchronized and collinearly superimposed with the FEL pulses. The FEL was circularly polarized and the probe laser pulses were polarized linearly along the He jet, that is perpendicularly to the spectrometer axis.
He droplets were formed in a supersonic expansion of cold He through an Even-Lavie-type pulsed valve. The mean number of He atoms per droplet, $\langle N\rangle$, was varied in the range $\langle N\rangle = 1\times 10^3$-$7\times 10^5$ by adjusting the temperature and opening time of the valve. The average number of excitations per droplet was controlled by $\langle N\rangle$ and the XUV intensity $I_\mathrm{XUV}$. The He jet intersected with the laser beams at right angles inside a magnetic-bottle spectrometer mounted at the Low-Density Matter (LDM) beamline perpendicular to the He jet~\cite{Lyamayev:2013}.

\begin{figure}
	\begin{center}
		\includegraphics[width=\columnwidth]{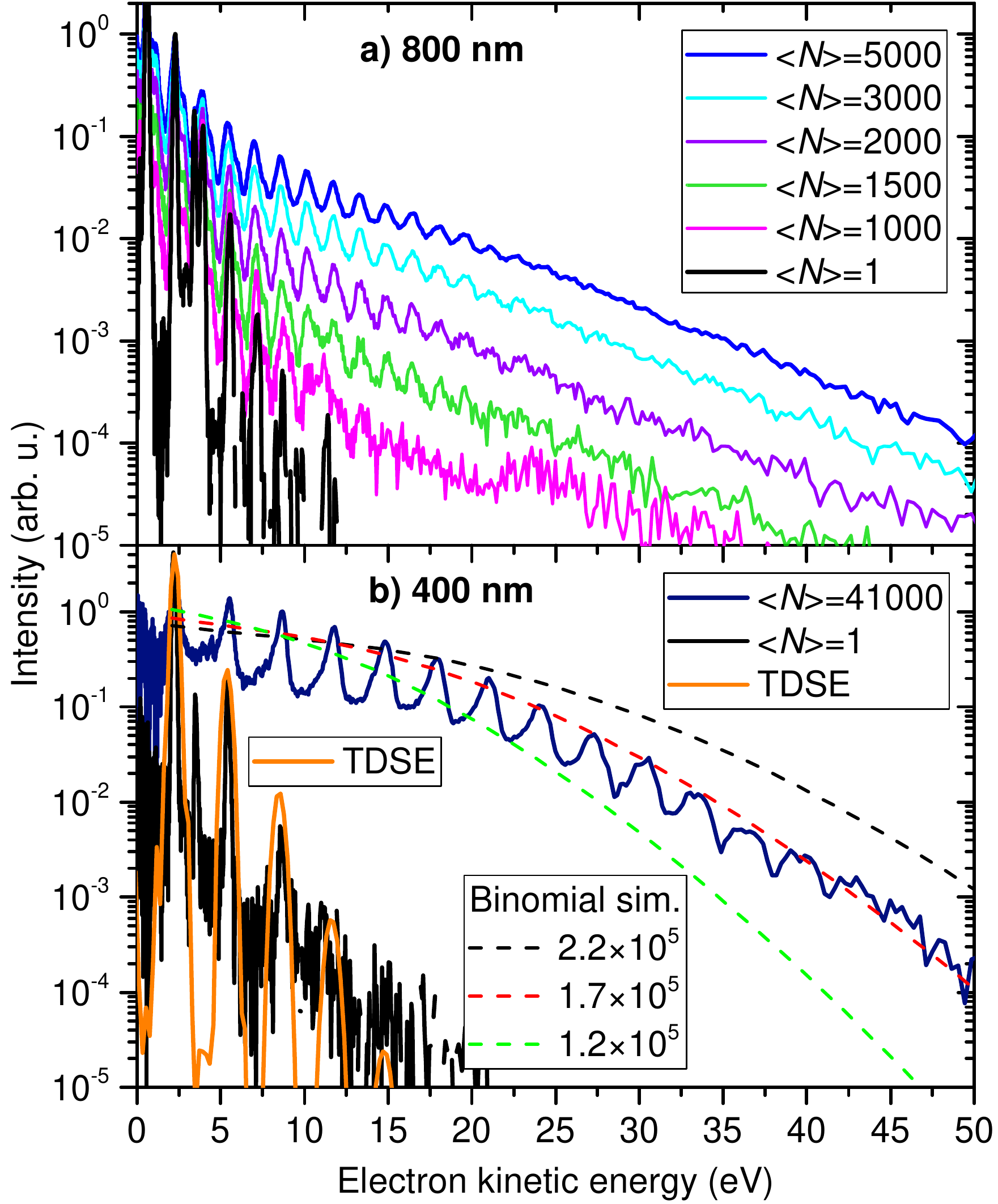}
		\caption{Photoelectron spectra of He droplets of different mean size $\langle N\rangle$ (colored lines) compared to He atoms (black lines, $\langle N \rangle = 1$). Both droplets and atoms are resonantly excited to the $1s4p$ state at a photon energy of 23.74~eV and ionized by NIR pulses a) and UV pulses b). The intensity of the probe pulses is $I_\mathrm{NIR,\, UV}\approx 10^{13}$~Wcm$^{-2}$. The two pulses were delayed by $\tau = 4~$ps in a) and temporally overlapped ($\tau = 0$) in b). See text for the discussion of the simulation curves (solid orange line for $N=1$ and dashed colored lines for different $N$).}
		\label{fig:VGL800400}
	\end{center}
\end{figure}

Fig.~\ref{fig:VGL800400} shows typical electron spectra of He atoms (black line) and He droplets (colored lines) that are resonantly excited to the $1s4p$ state and ionized by a probe pulse at the fundamental [800~nm, a)] or the second harmonic [400~nm, b)] of a Ti:Sa laser synchronized with the FEL. We observe massive enhancement of the high-energy electron yield for He droplets compared to He atoms for both probe wavelengths. The enhancement is particularly pronounced at high $E_e$ where the spectra are dominated by high-order ATI. The atomic ATI spectra are characterized by an exponential decay with superimposed ATI peaks up to 11~eV (7 photons) for NIR probe pulses and up to 15~eV (5 photons) for UV pulses. In contrast, using droplets, ATI electrons with energies up to 150~eV are observed, see SM Figs.~1 and 2 in the supplementary material (SM)~\cite{SM}. 

The colored lines in Fig.~\ref{fig:VGL800400}~a) show ATI spectra recorded for different $\langle N \rangle$. At $\langle N \rangle = 1000$ (magenta curve), ATI is only weakly enhanced compared to He atoms. For larger $\langle N \rangle$, the intensity of high-order ATI continuously increases and a plateau forms at $E_e\lesssim 25~$eV followed by an exponential drop towards higher $E_e$, see also SM Figs.~1 and 2~\cite{SM}.
Fig.~\ref{fig:VGL800400}~b) includes the result of a time-dependent Schr\"odinger equation (TDSE) calculation for He atoms in the $1s4p$ state (orange line). For details on the TDSE approach, see Ref.~\cite{kjeldsen2007solving} and the SM~\cite{SM}. The good agreement between the calculation and the experiment confirms the experimental determination of the probe pulse intensity. The peak at 3.5~eV in Fig.~\ref{fig:VGL800400} is due to ionization of the residual gas. Electron spectra measured for variable pump and probe pulse intensities are shown in SM Figs. 1 and 2, respectively~\cite{SM}.

\begin{figure}
	\begin{center}
		\includegraphics[width=\columnwidth]{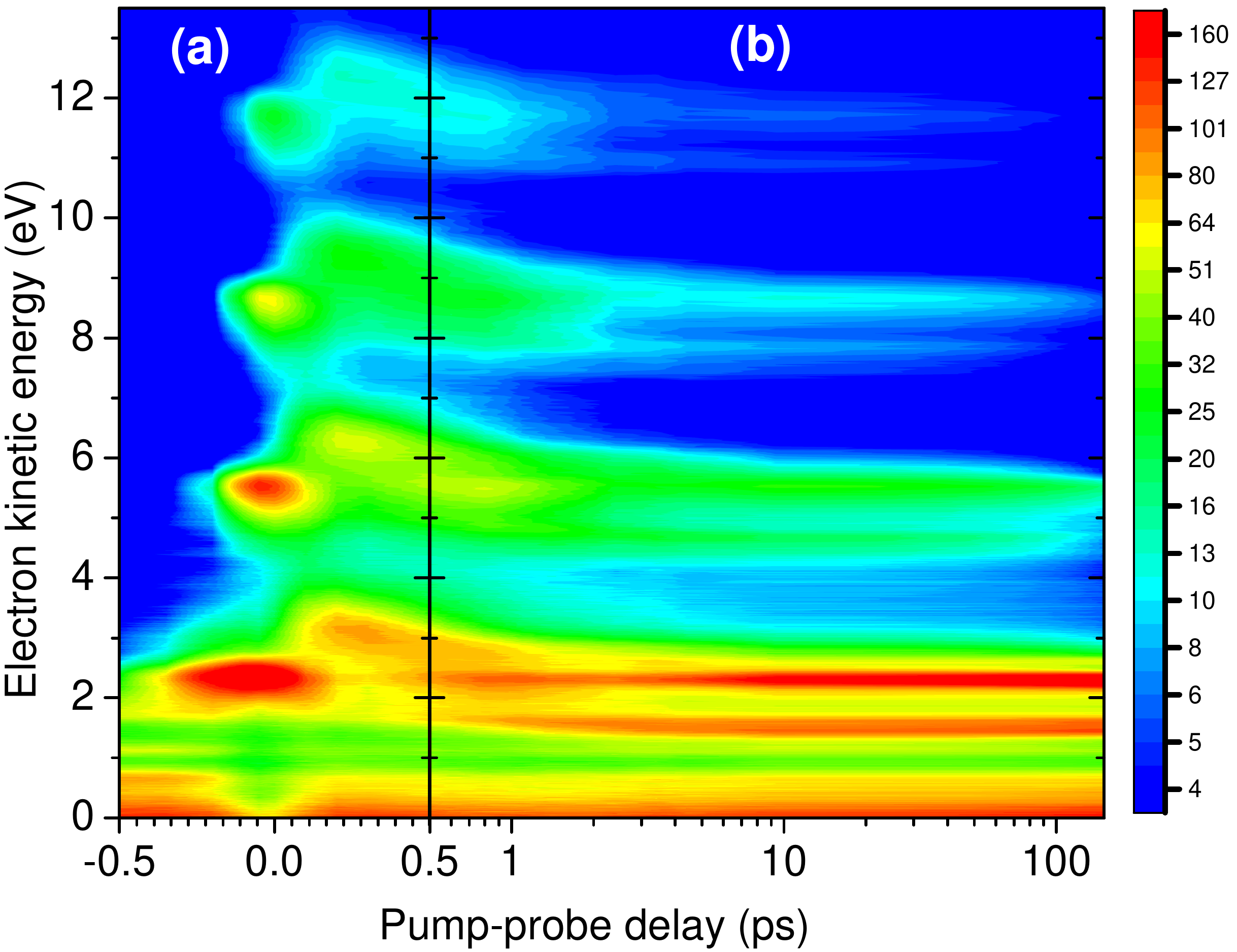}
		\caption{Logarithmic intensity plot of the electron spectra (vertical axis) as a function of the delay between the XUV pump and the UV probe pulses (horizontal axis). Red regions indicate high signal intensities, whereas the blue area shows the low-signal background. The short-time dynamics around zero delay is displayed in a) on a linear scale, the long-time dynamics is shown on a logarithmic delay scale in b). The mean droplet size is $\langle N \rangle = 41000$.}
		\label{fig:2D}
	\end{center}
\end{figure}
 
The time evolution of the ATI spectra due to the intrinsic relaxation dynamics of He nanodroplets~\cite{MudrichRelaxation,asmussen2021unravelling} is best seen for the UV probe pulses at low intensity $I_\mathrm{UV} = 10^{12}$~Wcm$^{-2}$ where only a few well separated ATI peaks are present. Fig.~\ref{fig:2D} shows the evolution of electron spectra as a function of the pump-probe delay $\tau$ around the temporal overlap of the two laser pulses, $\tau=0$, a), and for large $\tau$, b). At $\tau =0$, bright spots are observed at $E_e=2.3$~eV due to 1+1' resonant pump-probe ionization of the $1s4p$-excited He droplets and at $E_e=5.4$, $8.4$, and $11.6~$eV due to ATI up to $3^{rd}$ order. At $\tau\gtrsim 0.2~$ps, all ATI components spectrally broaden and slightly shift towards higher $E_e$. These delay-dependent changes directly reflect the dynamics of internal relaxation of the excited He nanodroplets~\cite{Ziemkiewicz:2015,MudrichRelaxation,asmussen2021unravelling}. For longer delays [Fig.~\ref{fig:2D} b)], the low-order ATI features split in two spectral components and  the higher-order ATI lines rapidly fade away.

\begin{figure}
	\begin{center}
		\includegraphics[width=\columnwidth]{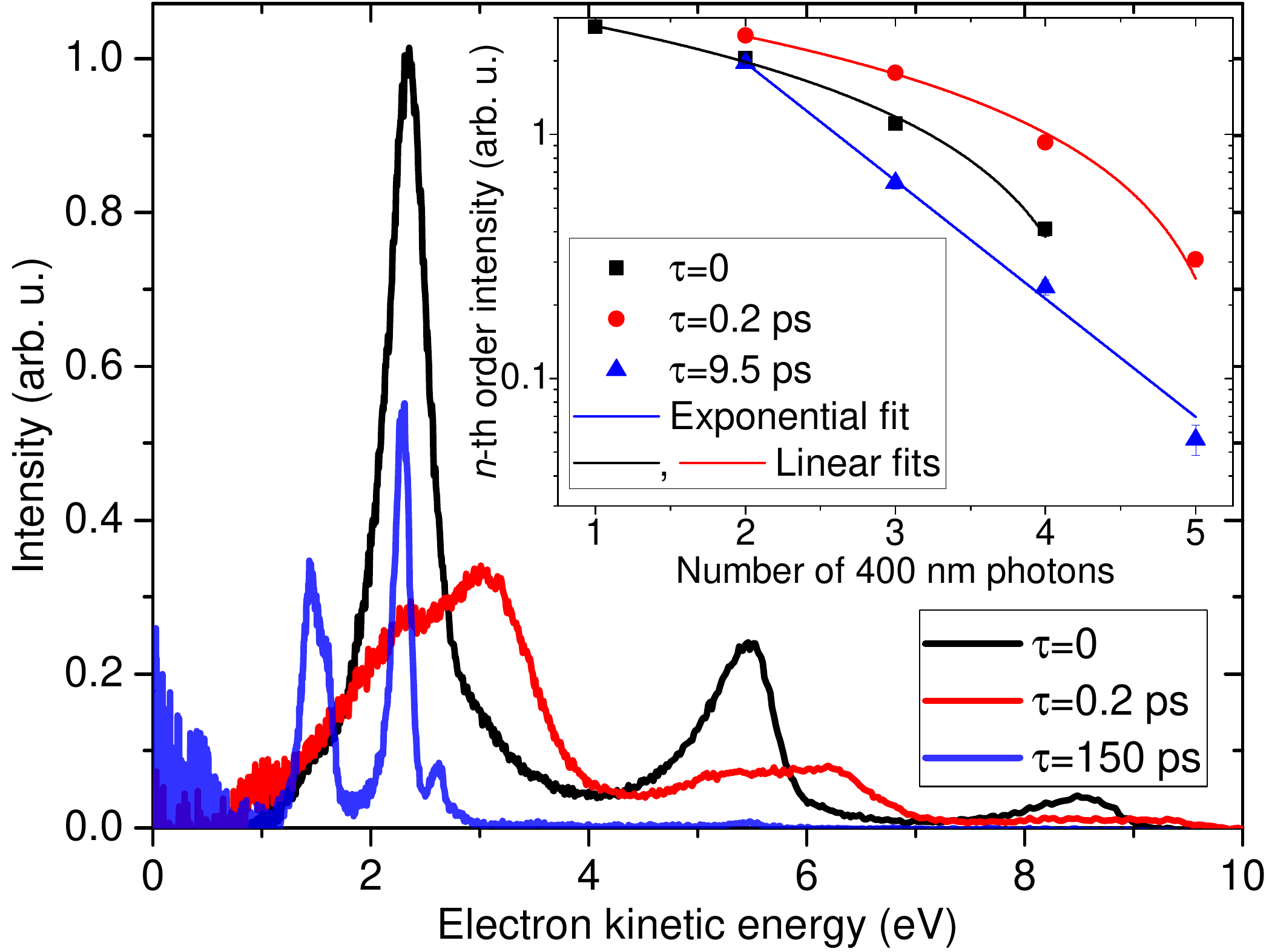}
		\caption{Electron spectra for UV probe pulses at pump-probe delays 0, 0.2, 9.5~ps (black, red, blue lines, respectively). The inset show on logarithmic scale the area of the ATI peaks including linear and exponential fits. The mean droplet size is $\langle N \rangle = 41000$. }
		\label{fig:ATIRelax}
	\end{center}
\end{figure}
For a more detailed analysis, we inspect the electron spectra for selected delays $\tau$, see Fig.~\ref{fig:ATIRelax}. These spectra were recorded at even lower probe pulse intensity $I_\mathrm{UV} = 2.5\times 10^{11}$~Wcm$^{-2}$ to allow a clear identification of the individual components. The three peaks at $\tau = 0$ (black line) exactly match the energies expected for direct ionization of the $1s4p$-excited He droplets by $n_{ph}=1$-$3$ probe photons, $E_e=2.3$, $5.4$, $8.4$~eV. At $\tau =200$~fs (red line), more than half of the population has relaxed into the $1s2s/2p$ droplet state as seen from the doubling of the peaks. 
The additional peaks are shifted up in energy by 0.4~eV because the $1s2s/2p$ droplet state is now ionized by $n_{ph}=2$-$4$ probe photons. At long delay ($\tau =150$~ps), the ATI peaks nearly vanish and the direct photoionization line splits in two ($E_e=1.6$ and $2.2~$eV) due to further relaxation into the two $^1S$ and $^3S$ spin components of the $1s2s$ He atomic state. The small peak at $E_e=2.6~$eV indicates the partial population of the $1s2p\,^3P$ state. Coincidentally, the lowest-order photoline of the final $1s2s\,^1S$ state at $E_e=2.2$~eV (two-photon ionization) nearly matches the one-photon $1s4p$ photoline at $\tau=0$ ($E_e=2.3$~eV). Note that the relaxed $1s2s/2p$-droplet peaks at $\tau =200$~fs ($E_e=3.2$, $6.3$, and $9.4$~eV) still show strong ATI. Thus, the drop of the ATI intensity at $\tau\geq 200$~fs cannot be ascribed to the increase of the number of probe photons needed for ionization. It is likely due to another aspect of the relaxation of excited He droplets: Following the localization of the droplet excitation on excited atoms~\cite{Closser:2014}, the electronic relaxation is accompanied by the decoupling of He$^*$ from the droplets due to the formation of void bubbles that eventually burst at the droplet surface, thereby releasing He$^*$ into vacuum~\cite{MudrichRelaxation,asmussen2021unravelling}. Clearly, ATI is enhanced in the early phase ($\tau\approx 0$) when $4p$ excitations are coupled to the droplet and likely delocalized, whereas ATI peaks fade away at the later stage ($\tau\geq 200~$fs) when the He$^*$ relax and detach from the droplet.

The ATI peak areas are shown in the inset of Fig.~\ref{fig:ATIRelax} for the spectra at $\tau = 0$, $200~$fs, and $9.5~$ps as a function of the number of absorbed probe photons, $n_{ph}$. As a characteristic feature, He droplet-enhanced higher order ATI peaks ($\tau=0$, $0.2$~ps) fall off more slowly compared to the ATI spectrum of the detached He$^*$ ($\tau =9.5$~ps), c.~f. Fig.~\ref{fig:VGL800400}. The He atomic ATI intensity drops exponentially as a function of $n_{ph}$, whereas the droplet-enhanced ATI clearly deviates from a pure exponential decay by forming a plateau-like structure followed by an exponential drop towards higher $E_e$.

\begin{figure}
	\begin{center}
		\includegraphics[width=0.95\columnwidth]{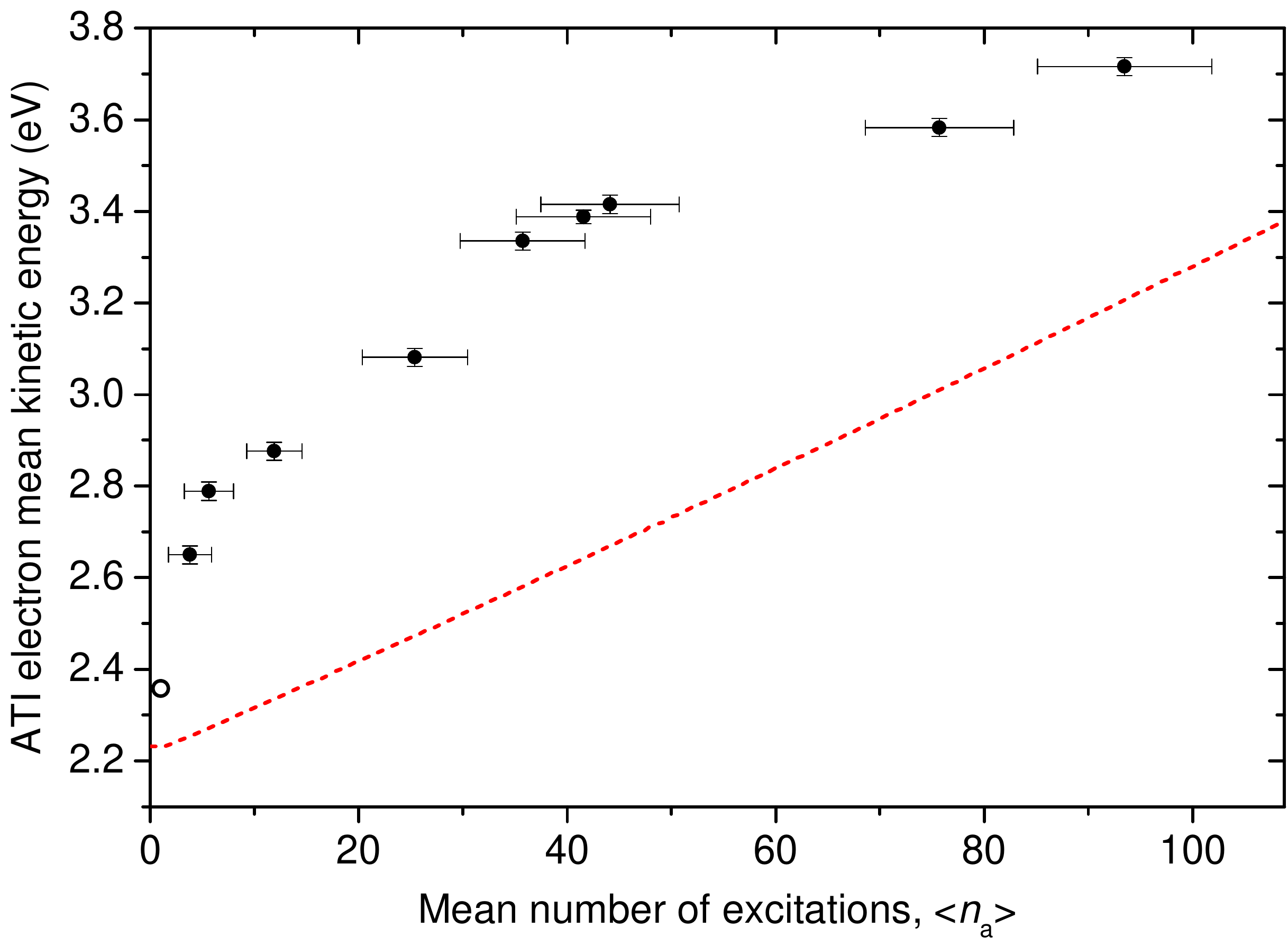}
		\caption{Mean electron kinetic energy as a function of the mean number of excitations per He droplet for
		$I_\mathrm{UV}=2.5\times 10^{11}$~Wcm$^{-2}$, see text. The dashed line is the result of the semi-empirical model fitted to the TDSE simulation for He atoms at that value of $I_\mathrm{UV}$. The open circle indicates the measured value for atomic He ($I_\mathrm{UV}=10^{13}$~Wcm$^{-2}$).
	}
		\label{fig:RelExc}
	\end{center}
\end{figure}

To quantify the He droplet-induced enhancement of ATI peaks independently of changes in the density of the He droplet jet, we consider the mean kinetic energy of the photoelectrons generated by the UV probe pulses, $\langle E_e \rangle =\int E_e\times S(E_e)dE_e/\int S(E_e)dE_e$, where $S(E_e)$ denotes the measured electron spectrum. In the atomic spectrum [Fig.~\ref{fig:VGL800400} b)], the ATI contribution is small, and $\langle E_e \rangle $ nearly equals the energy of the 1+2' peak, $E_e=2.23$~eV. For He droplet-enhanced ATI, we find that $\langle E_e \rangle $ strongly depends on both $\langle N \rangle $ and $I_\mathrm{XUV}$, see Fig.~\ref{fig:VGL800400} and SM Fig.~1. 
Therefore we consider the dependence of $\langle E_e \rangle $ on the mean number of excitations per droplet $\langle n_{\mathrm{a}} \rangle = \langle N \rangle\phi_\mathrm{XUV}\sigma_{2s \rightarrow 1s4p}$, shown in Fig.~\ref{fig:RelExc}. Here, $\phi_\mathrm{XUV}\propto I_\mathrm{XUV}$ is the XUV photon flux per pulse and $\sigma_{1s^2 \rightarrow 1s4p}$ is the resonant absorption cross section of the $1s4p$ state. The data points are inferred from measurements performed at $I_\mathrm{XUV}=4\times 10^8$ - $1.8\times 10^{10}$~Wcm$^{-2}$ and $\langle N \rangle =5\times 10^4$ - $1.8\times 10^6$. The smooth trend in the data confirms that $\langle n_\mathrm{a} \rangle$ is the characteristic parameter determining the ATI enhancement. 



A theoretical prediction of ATI in excited He nanodroplets within the existing frameworks is difficult. We observe ATI enhancement in the weak-field regime, where the quiver amplitude is small ($x_0<1~$\AA) and $U_p\sim10$~meV, as well as in the strong field regime, where $x_0\sim 5~$\AA~is comparable to the mean distance between the He$^*$ in one droplet, $d_\mathrm{He*}$, and $U_p\sim 1$~eV. 
Additionally, an accurate model would require the knowledge of excited state wave functions of the He droplet. 
A description extending of the three-step model to account for multiple scattering centers~\cite{Moreno:1994,Bandrauk:1997,Veniard:1999,Chirilua:2006,wang2020universal} is not applicable to our case, as even at moderate intensities ($10^{12}$~Wcm$^{-2}$) we see enhanced ATI, although $x_0\ll d_\mathrm{He*}$. Plasmonic enhancement by the collective coupling of the excitations~\cite{Tisch:1997,Zherebtsov:2011,Passig:2017} appears better suited here, as the He$^*$ in the droplet are highly polarizable~\cite{yan2000polarizabilities}.
However, the polarizability of the $1s2s$ state is more than two orders of magnitude smaller than that of the $1s4p$ state. In Figs.~\ref{fig:2D} and \ref{fig:ATIRelax} we observe only minor changes of the ATI enhancement when the $1s4p$ relaxes into the $n=2$ states, thus making a crucial influence of the droplet polarization on the ATI enhancement unlikely. Laser-assisted electron scattering, as observed in doped He droplets~\cite{treiber2021observation}, may play a role. However, the clear dependence on the number of active centers (Fig.~\ref{fig:RelExc} and SM Fig.~1) implies a different mechanism.

The model we propose here is based on the idea that all He$^*$ in a droplet collectively absorb photons from the probe pulse, and the total energy is channeled to a single He$^*$ which then emits an electron. Thus, the $n^{th}$ order of ATI is enhanced by the number of combinations resulting in the absorption of $n_{ph}+1$ photons by an ensemble of $n_a$ excitations, given by the binomial coefficient
\begin{equation}
{n_{ph}+n_a\choose n_a-1}.
\label{chap:ATI:eq_bin}
\end{equation}
Further details are given in the SM~\cite{SM}. Shortcomings of this model are, besides its simplistic assumption of energy transfer to one electron, the neglect of the He$^*$ relaxation dynamics, the broad distribution of He droplet sizes for a given $\langle N\rangle$, and the depletion of He$^*$ due to autoionization processes~\cite{Ovcharenko:2019,Laforge:2021}. Nevertheless, the resulting spectra reproduce the experimental data rather well, see the dashed red line in Fig.~\ref{fig:VGL800400} b) and in SM Fig.~2. The formation of a plateau at low $E_e$ followed by an exponential decay, as well as the shift of the cutoff towards higher $E_e$ for increasing $\langle n_{\mathrm{a}}\rangle$ is well reproduced [SM Fig.~5]. The dashed line in Fig.~\ref{fig:RelExc} shows the mean kinetic energy predicted by the model where the only adjustable parameter, an exponential decay constant obtained from a fit of the TDSE calculation for atomic He, is held constant. While the model fails to accurately match the experimental data, the overall rise of $\langle E_e \rangle $ with increasing  $\langle n_a\rangle$ is reproduced. Clearly, a more rigorous theoretical treatment is needed to accurately describe the laser-driven, collective dynamics of excited states in a nanometer-sized droplet.

In conclusion, resonant multiple excitation of He nanodroplets is shown to be a route to enhancing high-order ATI far beyond the atomic ATI cutoff. The enhancement can be controlled by the number of excitations per droplet, which is determined by the droplet size and the intensity of the XUV pulse. It is limited by the relaxation of excited states which evolve into free atoms~\cite{MudrichRelaxation,asmussen2021unravelling}, as well as by autoionization~\cite{Ovcharenko:2019,Laforge:2021}. 
Thus, the enhancement might be even more efficient for shorter pulses. Moreover, ultrashort ($\sim10~$fs), wavelength-tunable pulses would allow us to probe the role of the delocalization of excitations in different states of the droplet~\cite{Closser:2014}. The present study should also be extended to measuring the emission of XUV radiation from resonantly excited He nanodroplets to assess the potential of this scheme for enhancing HHG. Enhanced ATI and possibly HHG may be observed for other types of clusters as well, which can be resonantly excited by conventional laser pulses~\cite{Serdobintsev:2018}.

Financial support by the Deutsche Forschungsgemeinschaft (DFG) within projects STI 125/19-2 and STI 125/22-2 (Priority Programme 1840 “QUTIF”) is gratefully acknowledged. M.M. acknowledges support by the Carlsberg Foundation and R.F. thanks the Swedish Research Council (VR) and the Knut and Alice Wallenberg Foundation for financial support.\\

%

\end{document}


\title{Supplementary Information\\
	Collective enhancement of above threshold ionization by resonantly excited helium nanodroplets}

	\author{R. Michiels \textit{et al.}}
	\date{\today}
	
	\maketitle

\renewcommand\refname{SM References}
\renewcommand{\figurename}{SM Fig.}

\section*{Supplementary Note 1: Experimental electron spectra for varying pump and probe laser intensities} In the main text, we focus on the collective effect of multiply excited He in droplets leading to enhanced ATI. The main finding is that the enhancement of high-order ATI electron emission results from the collective coupling of multiple excitations per droplet prepared by the pump pulse and occurs for a wide range of pump and probe pulse intensities.

\begin{figure}
	\begin{center}
		\includegraphics[width=0.7\columnwidth]{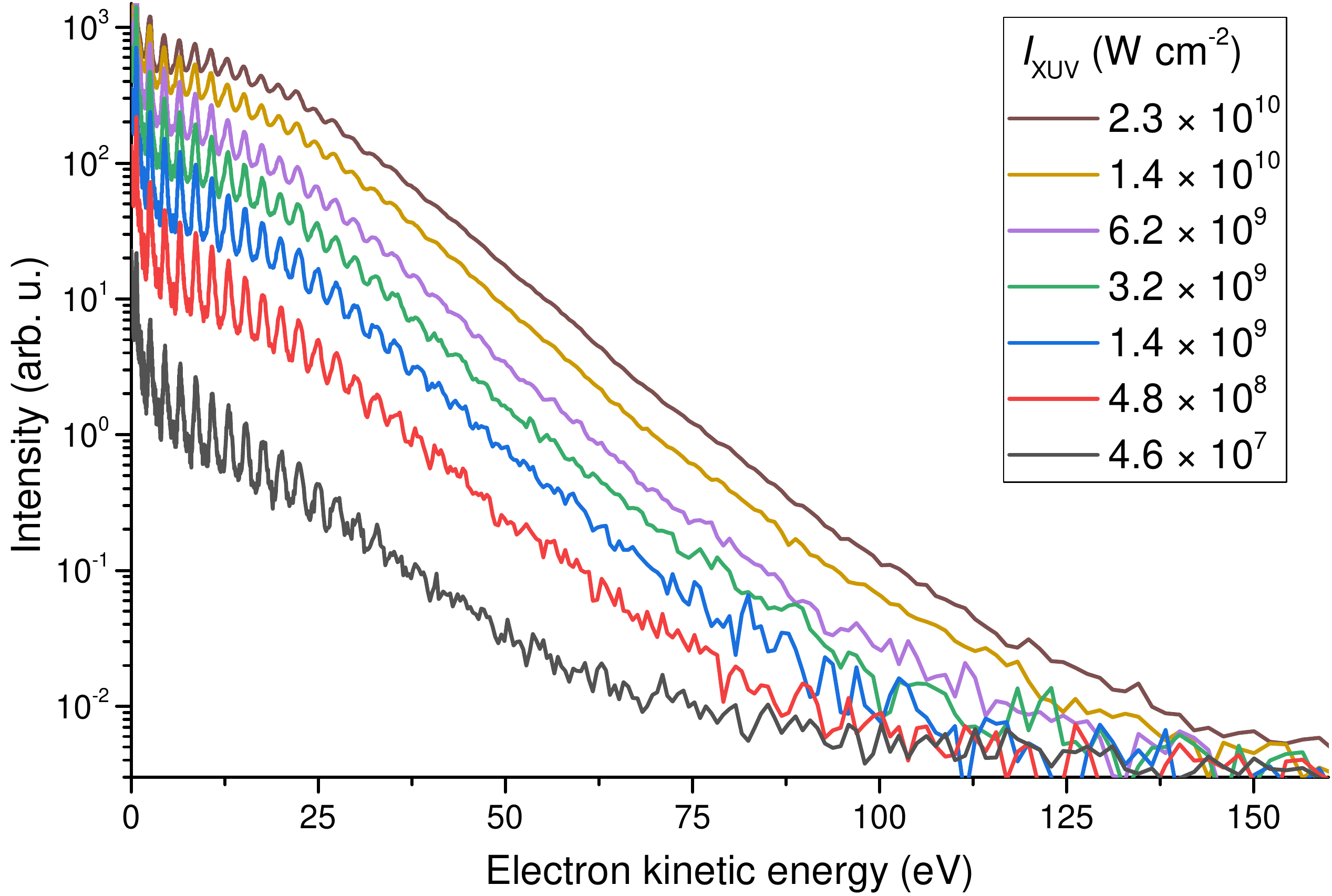}
		\caption{Experimental electron spectra for different XUV pump pulse intensities. The NIR probe pulses had an intensity of $I_\mathrm{NIR}=3.6\times10^{12}~$Wcm$^{-2}$. The mean droplet size was $\langle N\rangle=25000$. The spectra for atomic He (red lines) are scaled to roughly match the intensity of the first peak in the droplet spectra.} 
		\label{fig:ATI:SMFigXUV}
		\end{center}
\end{figure}

SM Fig.~\ref{fig:ATI:SMFigXUV} shows typical electron spectra recorded for variable the XUV pump pulse intensity. For increasing intensity, the spectra qualitatively change their structure, in that a plateau develops up to an electron energy of $E_e\approx25$~eV and the exponential drop shifts to high $E_e$ up to about 150~eV. 

\begin{figure}
	\begin{center}
		\includegraphics[width=\columnwidth]{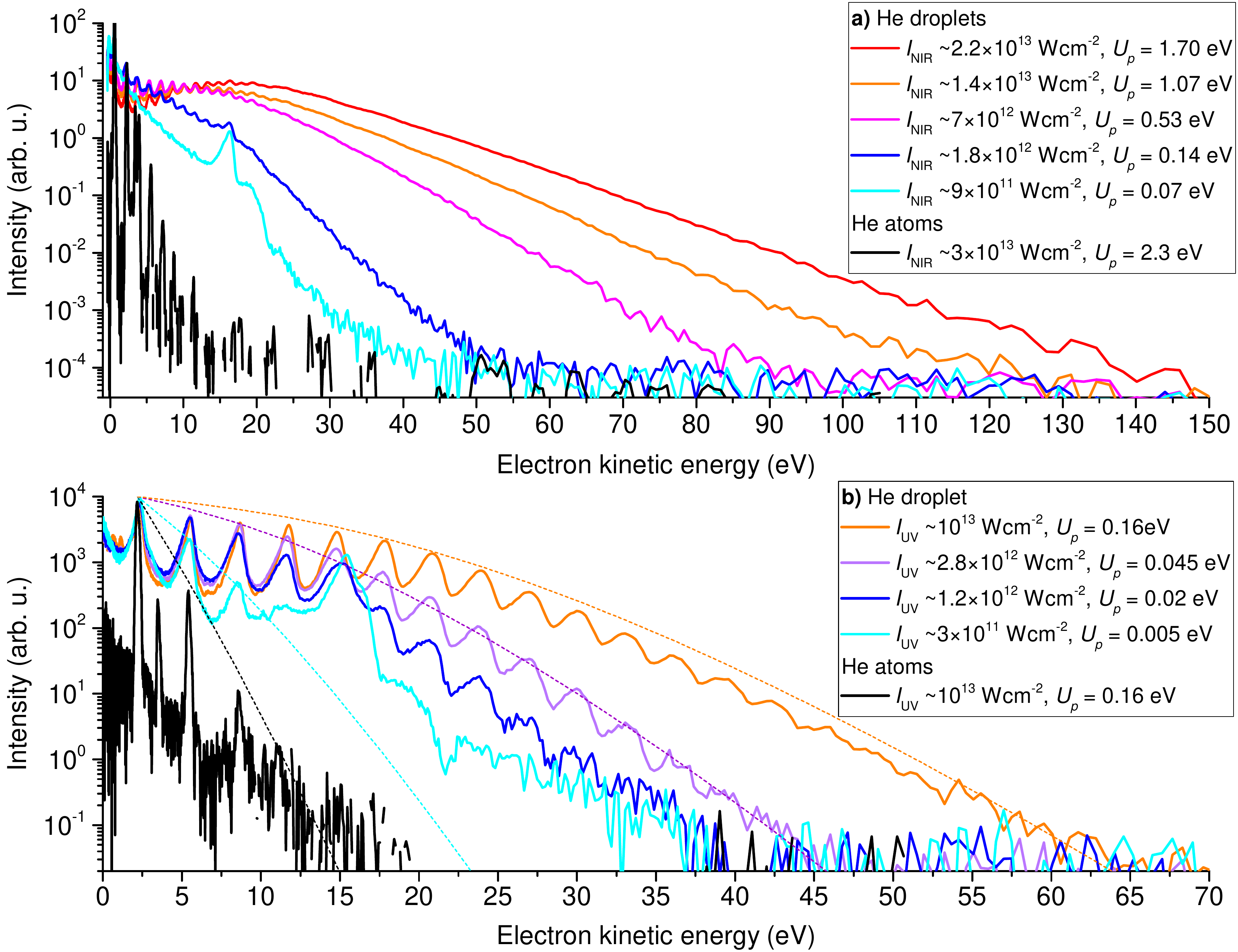}
		\caption{Electron spectra for different intensities of a) the 800-nm probe pulses, $I_\mathrm{NIR}$, and b) the 400-nm probe pulses, $I_\mathrm{UV}$ for an intensity of the XUV pulses of $I_\mathrm{XUV}=1.8\times10^{10}~$Wcm$^{-2}$. $U_p$ denotes the ponderomotive energy, see main text. The pump and probe pulses were delayed by 4~ps in a) and temporally overlapped in b). The mean He droplet size was $\langle N\rangle=25000$ in a) and $\langle N\rangle=41000$ in b). The smooth lines show the results of fitting the semi-empirical model [SM~Eq.\,(\ref{Eq:Final})] to the experimental data.}
		\label{fig:ATI:SMFig1}
	\end{center}
\end{figure}

SM Fig.~\ref{fig:ATI:SMFig1} shows electron spectra obtained for large He droplets ($\langle N\rangle=41000$) at variable intensity of the probe pulses [800 nm in a), 400 nm in b)]. The comparison with the atomic He ATI spectra obtained at the highest probe laser intensity highlights the ATI enhancement effect in He droplets. The peaks at 16~eV in the low-intensity spectra are due to autoionization of pairs of excited He atoms by interatomic Coulombic decay (ICD)~\cite{Ovcharenko:2019,Laforge:2021}.

\section*{Supplementary note 2: Computational details for the TDSE simulations}
\label{compdet}
The single-active-electron (SAE) potential describing the He atom was determined from quantum chemistry calculations, within the density functional theory, following the procedure in Ref.~\cite{hermann1988split}. We applied the split-operator spectral method of Hermann and Fleck~\cite{abu2010single} to obtain the energy and wavefunction corresponding to the He~4p state. We started the time propagation from an inital guess wavefunction, denoted by $\psi(\bm{r},t=0)$ and used the SAE potential for He. For the inital state, we chose the hydrogenic 2p$_x$ state for convenience. We performed field-free propagation for a time duration of 1000 a.u. and saved the wavepacket every 1 a.u. of time. From the time-dependent wavefunction obtained in this manner, $\psi(\bm{r},t)$, an autocorrelation function $\mathcal{P}(t)= \langle\psi(\bm{r},t=0)|\psi(\bm{r},t) \rangle$ was calculated, and a bound state spectrum $\mathcal{P}(E)$ was obtained as 
\begin{equation}
\mathcal{P}(E) =     \frac{1}{T}\int_{0}^{T}dt~w(t)~\exp(iEt)~\mathcal{P}(t),
\end{equation}
where $w(t)$ is the Hanning window function~\cite{abu2010single}. From the bound state spectrum, the energy of the 4p excited state was -0.0202~a.u., compared with -0.0323 from the NIST database for atomic levels. Once the He~4p energy was well resolved in the $\mathcal{P}(E)$ spectrum, the corresponding wavefunction was constructed and normalized as follows
\begin{equation}
\label{wfc}
\psi(\vec{r}) = \frac{1}{T}\int_{0}^{T}dt~w(t)~\exp(iEt )\Psi(\vec{r},t).
\end{equation}
We used the wavefunction of SM~Eq.~(\ref{wfc}) as the initial state for the solution of the TDSE, after we had tested for numerical stability of the wavefunction on the radial grid.

In our TDSE method, the time-dependent wavefunction $\psi(\textbf{r},t)$ is represented by a partial wave expansion in which the spherical harmonics $Y_{lm}(\Omega)$ are used to describe the angular degrees of freedom and a radial grid is used for the time-dependent reduced radial wave functions, $f_{lm}(r,t)$, i.e.,  
\begin{equation}
    \label{Eq1}
\psi(\textbf{r},t)=\sum_{lm} \frac{f_{lm}(r,t)}{r} Y_{lm}(\Omega). 
\end{equation}

The TDSE is solved for the electron in the SAE potential discussed above in the presence of the external field. The TDSE is propagated in the velocity gauge (VG)~\cite{kjeldsen2007solving} with a combined split-operator~\cite{abu2010single} Crank-Nicolson method. The electric field $E(t)$, linearly-polarized along the $z$-axis, is defined as
\begin{equation}
    \vec{E}(t) = -\partial_t A(t) \hat{z}= -\partial_t \left(\frac{E_0}{\omega}\sin^2(\pi t/\tau)\cos(\omega t+\phi) \right)\hat{z},
    \label{E_field}
\end{equation}
where $E_0$ is the field amplitude, $\omega$ the frequency, and $\phi$ the carrier-envelope phase (CEP) for a laser pulse with duration $\tau$. The TDSE calculations were performed at laser frequencies of $\omega=0.057$~a.u. and 0.114~a.u. corresponding, respectively, to wavelengths of 800~nm and 400~nm. The CEP is kept fixed  ($\phi=-\pi/2$), the pulses contained 10 optical cycles, and the laser intensities 1$\times10^{13}$~Wcm$^{-2}$ at 400~nm and $5\times10^{13}$~Wcm$^{-2}$ at 800~nm.

In the TDSE calculations for He, the radial grid contains 8192 points and extends to 800 a.u. The size of the angular basis set is limited by setting $l_{max}=60$ in SM~Eq.~(\ref{Eq1}). The ATI spectra were produced by projecting the wavepacket at the end of the laser pulse on scattering states of the He atom, produced using the SAE potential. 

\begin{figure}
	\begin{center}
		\includegraphics[width=0.7\columnwidth]{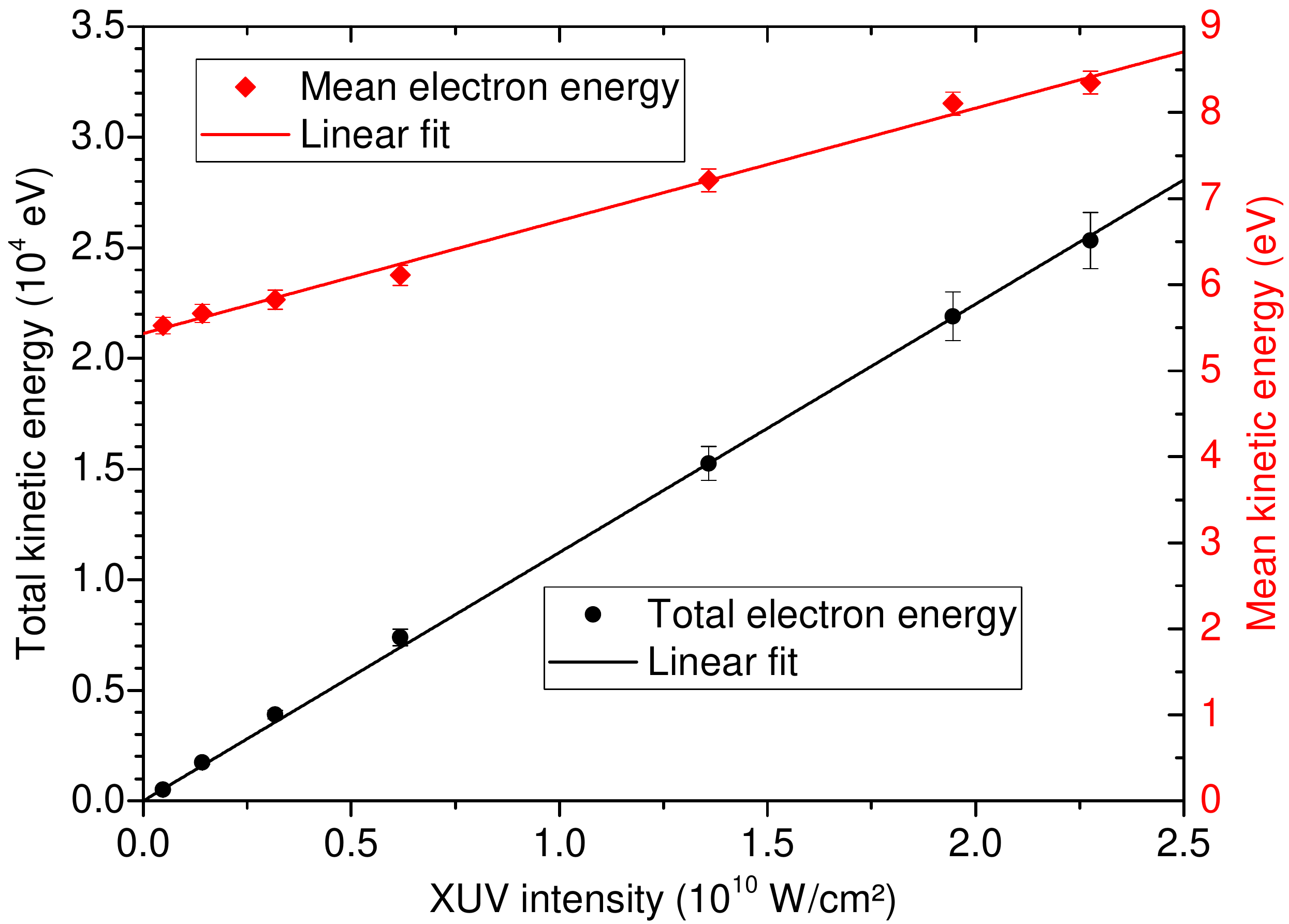}
		\caption{Total kinetic energy according to SM~Eq.~(\ref{eq:totalE}) based on the measured energies and numbers of electrons per laser shot (black squares) and mean electron energy (SM~Eq.~(\ref{eq:meanE})) (red squares) for He droplets of size $\langle N\rangle=25000$. The intensity of the 800-nm-probe pulses was $I_\mathrm{NIR}=3.6\times10^{12}~$Wcm$^{-2}$.} 
		\label{fig:ATI:SMFig2}
	\end{center}
\end{figure}

\section*{Supplementary Note 3: Detailed description of the semi-empirical simulation} The basic idea of the proposed model [Eq.~(1) in the main text] is that an ensemble of $n_a$ excited atoms, He$^*$, in the droplet collectively absorbs $n_{ph}$ photons from the probe pulse. The absorbed energy is then channeled towards a single He$^*$ which emits an energetic electron. The emission of an electron of the $n^{th}$ ATI order is enhanced by the number of combinations of $n_a$ He$^*$ absorbing $n_{ph}$ photons. This factor is given by the binomial coefficient
\begin{equation}
	{n_{ph}+n_a-1\choose n_a-1},
	\label{chap:ATI:eq_bin}
\end{equation}
where $n_{ph}$ is related to the ATI order $n$ by $n=n_{ph} - 1$. 
For example, to generate an ATI electron of second order, three photons need to be absorbed by the ensemble. An ensemble containing two excited atoms has four possibilities of absorbing three photons: [3, 0], [2, 1], [1, 2] and [0, 3]. If the excited atoms are separated, only two of the combinations are possible: [3, 0] and [0, 3]. The resulting enhancement factor is 2.
To account for the probability of absorption of $n_{ph}$ photons by a free He$^*$ as observed in the experiment, the enhancement factor is additionally weighted by 
\begin{equation}
	P^{at}(n_{ph}) = P_0\exp(-cn_{ph}).
	\label{eq:expdecay}
\end{equation}
Here, we approximate the gas phase ATI intensity distribution by an exponential decay. 
The constant $c$, which defines the ATI order at which the electron intensity drops to $1/e$ of the first order, depends on the probe pulse intensity and can be obtained either from the TDSE simulation discussed in the supplementary note 2, or from fitting the experimental ATI spectrum for He atoms. Combining the two factors gives us the expected probability for absorbing $n_{ph}$ photons in an ensemble containing $n_a$ He$^*$,
\begin{equation}
	P_{drop}(n_{ph},n_a) = {n_{ph}+n_a-1\choose n_a-1}\,P_0 \exp(-c\,n_{ph}).
	\label{chap:ATI:eq_bin_2}
\end{equation}

\begin{figure}
		\begin{center}        	\includegraphics[width=0.7\columnwidth]{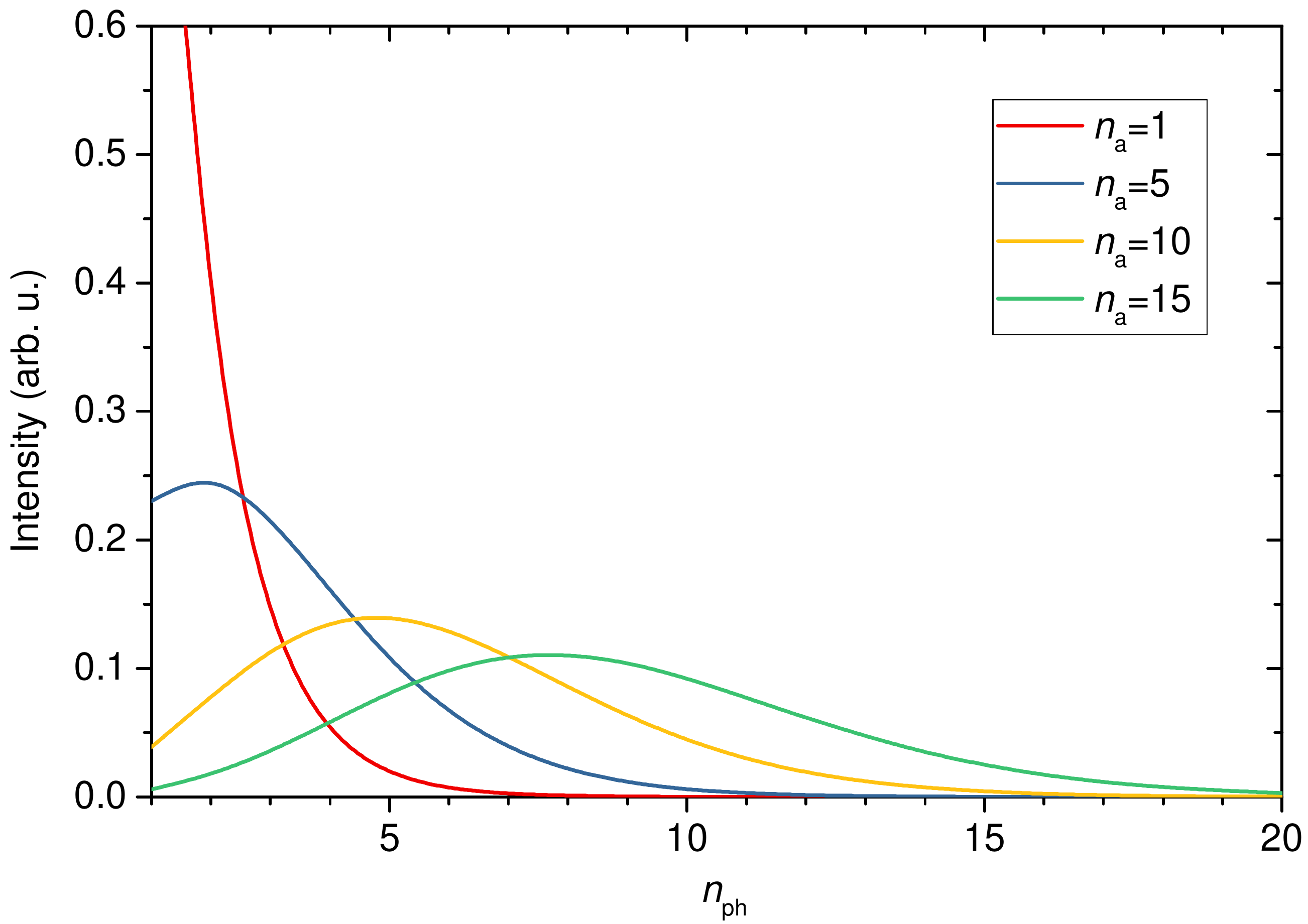}
			\caption{Simulated electron spectra as a function of the number of absorbed photons, $n_{ph}$, for different numbers of excited He atom per droplets, $n_a$, according to SM~Eq.~(\ref{Eq:ATINorm}).}
			\label{fig:ATI:SimSpec}
		\end{center}
\end{figure}

For a certain ATI order $n$, SM~Eq.~(\ref{chap:ATI:eq_bin_2}) predicts that the intensity increases according to factorial scaling as a function of $n_a$. This steep increase gives rise to an unphysical behavior in the limit of large $n_a$ to the extent that the number of absorbed photons diverges. A characteristic experimental finding is that the total energy of detected electrons, $E_{tot}$, rises proportionally to $n_a$, see SM Fig.~\ref{fig:ATI:SMFig2}. $E_{tot}$ per He$^*$ is approximately given by the total number of absorbed probe photons. These conditions are implemented into SM~Eq.~(\ref{chap:ATI:eq_bin_2}) by normalizing with respect to $n_a$,

\begin{equation}
	P_{norm}(n_{ph},n_a) = P_{drop}(n_{ph},n_a)\,\times \frac{n_a}{\sum_{n_{ph}}n_{ph}P_{drop}(n_{ph},n_a)}.
	\label{Eq:ATINorm}
\end{equation}
The resulting expected electron intensities calculated from SM~Eq.~(\ref{Eq:ATINorm}) and for different $n_a$ versus $n_{ph}$ are shown in SM Fig.~\ref{fig:ATI:SimSpec}. 

When converting the horizontal axis to $E_e$ by multiplication by probe photon energy, these distributions can be compared to measured electron spectra. As $n_a$ rises, the maximum electron intensity shifts towards larger $n_{ph}$ followed by an exponential drop towards higher $E_e$. These trends are in good agreement with the experimental spectra (SM Fig.~\ref{fig:ATI:SMFigXUV} and Fig.~1 of the main text). The model consistently connects to the atomic case in that for $n_a=1$ the exponentially decaying intensity measured for the He atomic beam is recovered. However, the model deviates from the experimental data in that the low-energy part of the spectra does not decrease as predicted by the model. This is related to the fact that the experimental data result from averaging over the intensity distribution in the focal volume of the pump laser and over the distribution of He droplet sizes. 

As the experimental conditions are well known, we can account for the focal volume averaging assuming Gaussian intensity profiles of the laser beams. The XUV focus was measured to have a FWHM of 70~$\mu$m, for the UV probe laser the FWHM spot size was 80~$\mu$m. The Rayleigh lengths of both laser foci are much larger than the size of the He jet. Thus, we assume a constant intensity in the propagation direction of the lasers. Furthermore, we assume cylindrical symmetry with respect to the beam axes. Focal volume averaging is implemented numerically by sampling the radius $R$ in 1-nm steps and thus obtaining a mean laser intensity for each radial sample. From this we calculate the mean number of He$^*$ per droplet per sample of the focal volume, $n_a(R)$. The intensity distribution of the probe laser is accounted for by introducing a correction factor in SM~Eq.~(\ref{Eq:ATINorm}) specifying the reduced probability of photon absorption in the outer region of the focus due to the lower laser intensity. The expected electron intensity for $n_{ph}$ absorbed photons in one sample with radius $R$ then reads
\begin{equation}
	I(n_{ph},n_a(R)) = P_{norm}(n_{ph},n_a(R))\, \Phi(R).
		\label{Eq:Final}
\end{equation}

	\begin{figure}
		\begin{center}			\includegraphics[width=0.7\columnwidth]{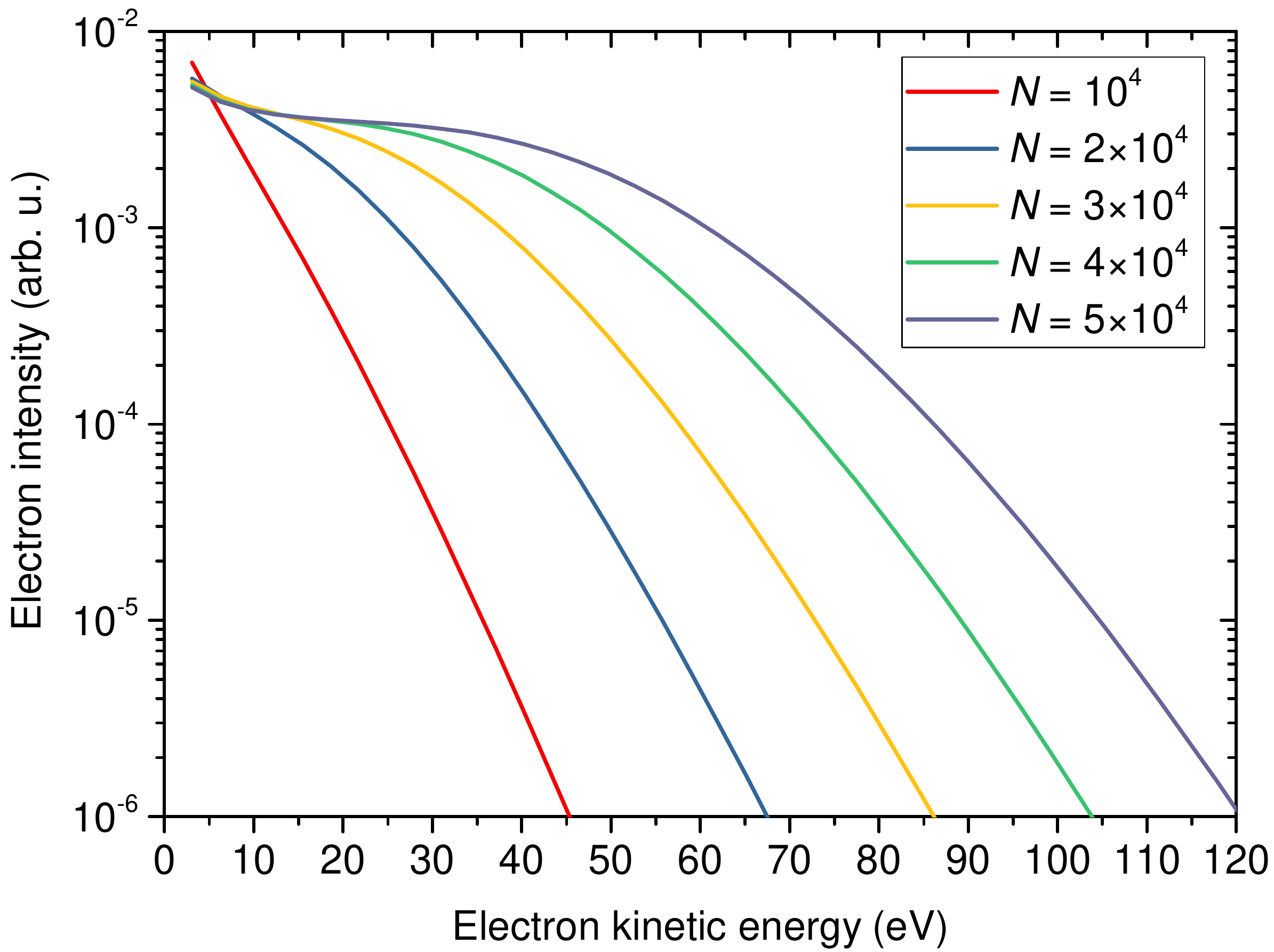}
			\caption{Focal volume-averaged electron spectra for different droplet sizes $N$ calculated using SM~Eq.\,(\ref{Eq:Final}).}
			\label{fig:ATI:SimSpec_FA}
		\end{center}
	\end{figure}
Here, $\Phi(R)$ is the radial distribution of the photon flux of the probe pulses. To obtain the final simulated electron spectra, we weight the contributions from each radial sample with the volume element and sum over the total focal volume. SM Fig.~\ref{fig:ATI:SimSpec_FA} shows the result for various droplet sizes and a constant exponential parameter $c$ in SM~Eq.~(\ref{eq:expdecay}). When using the mean droplet size $\langle N\rangle$ as a free fit parameter, we can fit our model to the experimental spectra, as shown in Fig.~1 b) of the main text for the experimental values $I_\mathrm{XUV}=1.8\times10^{10}~$Wcm$^{-2}$ and droplet sizes $N=1.2$, $1.7$, and $2.2\times 10^5$. The exponential decay constant was determined from the TDSE simulation. The simulated curves shown in SM Fig.~\ref{fig:ATI:SMFig1} b) are obtained for the experimental values of the droplet sizes while using the exponential constant as a fit parameter. The main features of the experimental spectra in Fig.~1 b) and in SM Fig.~\ref{fig:ATI:SMFig1} b) are well reproduced. The increasing intensity towards low $E_e$ is due to the high statistical weight of the low intensities in the extended outer regions of the focal volume. For increasing droplet size, that is for rising $n_a$, a plateau develops which extends towards higher kinetic energies, followed by an exponential decay. Accordingly, $\langle E_e\rangle$ rises for increasing $\langle n_a\rangle$, as seen in Fig.~4 of the main text, where the exponential decay constant was determined from a fit of the He atomic ATI spectrum from the TDSE calculation. We refrain from averaging over droplet sizes in this simulation. It can be estimated that the log-normal droplet-size distribution has a similar effect as the focal volume averaging, \textit{i.\,e.}, a mixing of ensembles with different $n_a$ leading to an overall broadening of the simulated spectrum.


\section*{Supplementary Note 4: Consistency of the model assumptions with the experiment} 
\begin{figure}
	\begin{center}
		\includegraphics[width=0.7\columnwidth]{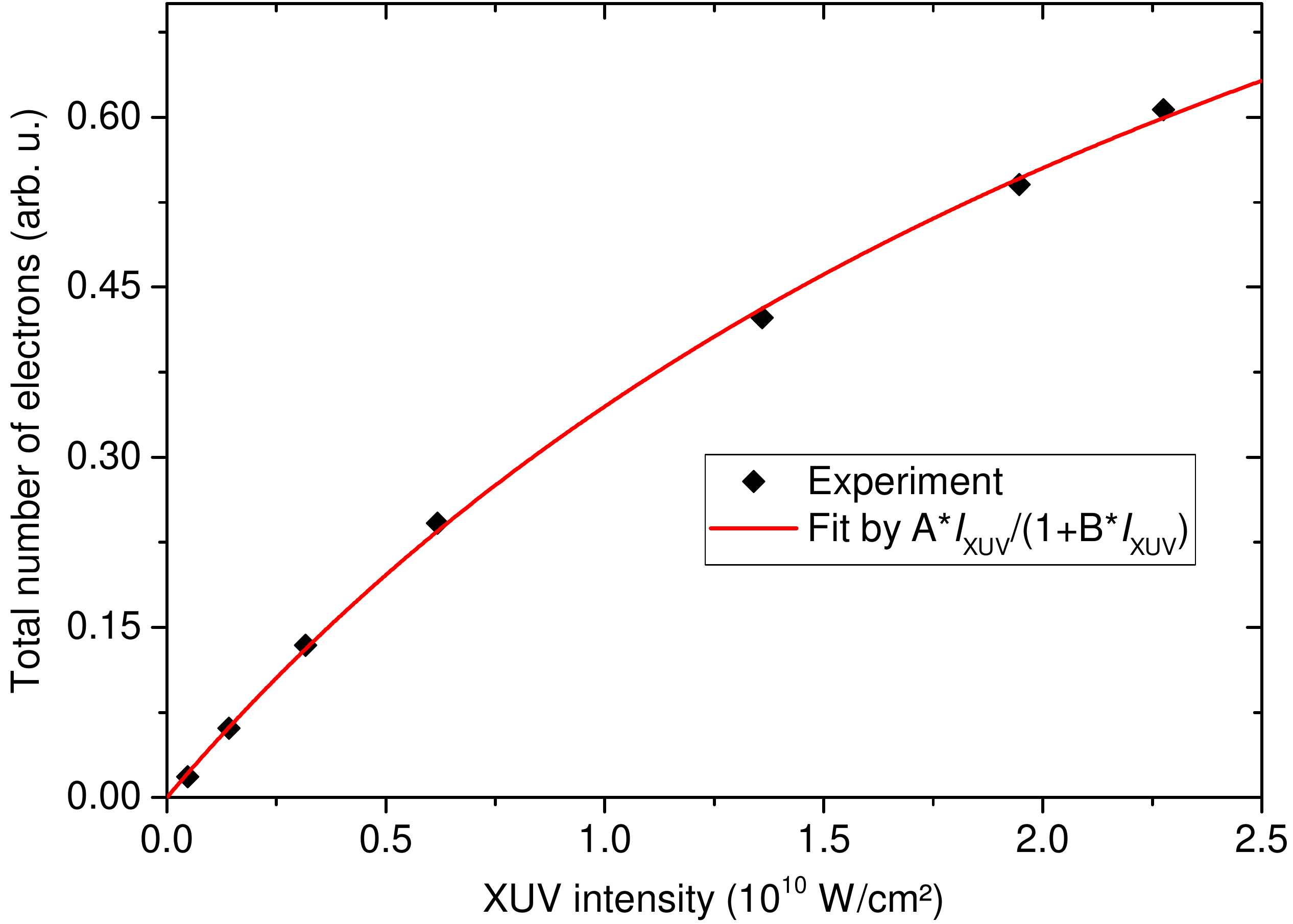}
		\caption{Total number of detected electrons as a function of XUV intensity. The NIR pulse intensity was $I_\mathrm{NIR}=3.6\times10^{12}~$Wcm$^{-2}$ and the mean droplet size was $\langle N\rangle=25000$.}
		\label{fig:ATI:convergingesum}
	\end{center}
\end{figure}

In our model, the energy absorbed by all He$^*$ is channeled towards one He$^*$ which emits an energetic electron. As a consequence, the remaining He$^*$ cannot be ionized, and the ionization probability per He$^*$ should decrease with increasing ATI enhancement. Confirming this conjecture by comparing with experimental data is an important consistency check. There are two ways of controlling the ATI enhancement: By varying the mean droplet size, $\langle N\rangle$, at constant XUV intensity $I_\mathrm{XUV}$ or by changing the number $n_a$ of excitations per droplet by varying $I_\mathrm{XUV}$. Since in the experiment, a change of $\langle N\rangle$ is intrinsically coupled to a change of target density, it is difficult to unambiguously probe the droplet-size dependence. Therefore we keep $\langle N\rangle$ fixed and vary $I_\mathrm{XUV}$, see SM Fig.~\ref{fig:ATI:convergingesum}. The total number of electrons indeed deviates from a linear increase with $I_\mathrm{XUV}$. It is well reproduced by a simple saturation model (smooth solid line), indicating that the probability for He$^*$ ionization decreases with rising ATI enhancement. 

To quantify the ATI-enhancement effect we consider the kinetic energy of electrons emitted by droplet-enhanced ATI as a function of $I_\mathrm{XUV}$. The mean electron kinetic energy
\begin{equation}
\langle E_e\rangle =\int E_e\times S(E_e)dE_e/\int S(E_e)dE_e,
\label{eq:meanE}
\end{equation}
shown as red squares in SM Fig.~\ref{fig:ATI:SMFig2}, increases nearly linearly with XUV intensity. Indeed, higher orders are generated for higher degrees of droplet excitation. This is a clear manifestation of ATI enhancement as discussed in the main text. Additionally, we observe that the total kinetic energy of all electrons in the experimental spectrum $S(E)$, 
\begin{equation}
E_{tot}=\int E_e\times S(E_e)dE_e    
\label{eq:totalE}
\end{equation}
(black squares), rises proportionally to the XUV intensity. As the He($1s^2\rightarrow 1s4p$) transition is not expected to be saturated at any XUV intensity used here, the total number of He$^*$ also increases proportionally to $I_\mathrm{XUV}$. With $I_\mathrm{XUV}=2.3\times 10^{10}$~Wcm$^{-2}$, 70~fs pulse duration (FWHM), and a cross section for resonant absorption at a photon energy of 23.7~eV of 2.9~Mb~\cite{Joppien:1993}, we expect roughly 0.1\% of the He atoms in the droplet to be excited. Thus, the total electron energy scales with the number $n_{a}$ of He$^*$ as $E_{tot}\propto n_{a}$, indicating that the number of photons absorbed per He$^*$ is nearly independent of the ATI enhancement effect.

%